\documentclass[a4paper,twocolumn]{esapub}

\renewcommand{\deg}{$^\circ$}

\usepackage{times, natbib, graphicx}
\title{HIGH ACCURACY MATCHING OF PLANETARY IMAGES}
\author{Giuseppe Vacanti}
\author{Ernst-Jan Buis}
\affil{cosine Science \& Computing BV, Niels Bohrweg 11, 2333 CA,
  Leiden, The Netherlands}
\begin{document}
\maketitle
\begin{abstract}
  We address the question of to what accuracy remote sensing images of
  the surface of planets can be matched, so that the possible
  displacement of features on the surface can be accurately measured.
  This is relevant in the context of the libration experiment aboard
  the European Space Agency's BepiColombo mission to Mercury.  We focus
  here only on the algorithmic aspects of the problem, and disregard
  all other sources of error (spacecraft position, calibration
  uncertainties, \textit{etc.}) that would have to be taken into
  account. We conclude that for a wide range of illumination
  conditions, translations between images can be recovered to about
  one tenth of a pixel \textit{r.m.s.}
\end{abstract}

\section{Introduction}
\label{sec:introduction}

One of the goals of the European Space Agency's BepiColombo mission
to Mercury is the measurement of the amplitude of the libration of
Mercury. In order to do this images of the same surface areas will be
taken at different times during the libration cycle and compared. When
all other effects---spacecraft position, Mercury's rotation,
spacecraft attitude, \textit{etc.}---are taken into account, any
remaining discrepancy between the positions of features on the surface
must be due to the libration of the crust of the planet.

Here we address the question of to what accuracy images can be
matched, and we focus only on the algorithmic aspects of the problem,
disregarding all other sources of error that would have to be taken
into account to solve the scientific problem. We shall show that by
using a shape-based matching algorithm images taken under a wide range
of illumination conditions can be matched to one tenth of a pixel
root-mean-square. Based on this we conclude that the accuracy of the
pattern matching algorithm is not the limiting factor in the ultimate
accuracy that can be achieved by the libration experiment on BepiColombo.

\section{Pattern Matching}
\label{sec:pattern-matching}

\begin{figure}[htbp]
  \centering
  \includegraphics[width=8cm]{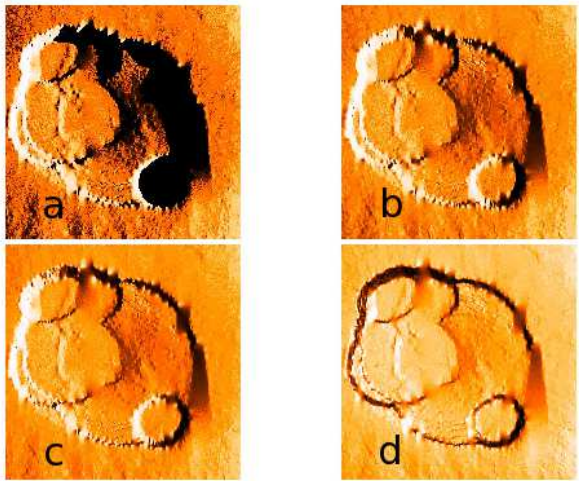}
  \caption{A digital elevation model of Olympus Mons viewed by an
    imaging camera under different illumination conditions: (a) The Sun is at $\mathit{5}^\circ$
    elevation; (b) The Sun is at $\mathit{25}^\circ$ elevation; (c) The Sun is at
    $\mathit{50}^\circ$ elevation; (d) The Sun is at $\mathit{85}^\circ$ elevation. In all
    cases the Sun's azimuth is $\mathit{0}^\circ$ (to the right).}
  \label{fig:sample-images}
\end{figure}

The pattern matching algorithms to be used in this study will have to
deal with images that may appear to be drastically different from one
another, still they refer to the same region. Consider for example the
images shown in figure~\ref{fig:sample-images}. To the human eye it is
clear that the images refer to the same region, but any algorithm that
relied on the presence of identical features in the images would have
great difficulty concluding that the images are related at all.

What is clear by visual inspection is that a number of edges---sharp
changes in the level of illumination between contiguous pixels---are
common between images. These edges appear where sharp changes in the
altimetric profile occur. It is also clear that not all edges appear
in all images, owing to the complex interplay between the position of
the Sun, and the orientation and slope of the features on the ground.

Compare for instance images \textit{b} and \textit{d} in
figure~\ref{fig:sample-images}: the left rim of the crater is bright
in one image, and dark in the other. No similar change is observed on
the right rim of the crater.

Take now images \textit{a} and \textit{c}. Here the left rim of the
crater appears almost to be the same, but the extent of the shadow
cast by the right rim is dramatically different.

The ideal algorithm must be able to identify the edges in the two
images, must be robust against local, non-linear changes in
illumination conditions, and it must be able to operate by identifying
a subset of features that are common to the pair of images being
compared. Finally, based on the common features identified, the
algorithm must be able to recover a possible translation between the
two images.

Algorithms that try to minimize the difference between the two,
possibly scaled, images are clearly not going to be suited for the
task, unless the images to be compared are taken under very similar
illumination conditions. While this is possible, it would be a very strong
constraint on the operations of a mission.

Based on the considerations above, we have chosen to make use of the
image matching algorithms available in the HALCON software library
\cite{halcon}. This is a commercial product used in image vision and
image recognition applications.

One particular technique available in the HALCON library is the
so-called \emph{shape-based matching} \cite{halcon_manual}. This
technique is based on an algorithm that identifies the shape of
patterns in images, and can be instructed to find in a comparison
image the shape identified in a reference image.

\subsection{Shape-Based Pattern Matching}
\label{sec:shape-based-pattern}

The detailed description of the algorithm can be found in the HALCON
documentation \cite{halcon_manual} and has been submitted as part of a
European Patent Application~\cite{espacenet}.

The algorithm proceeds through the following steps:

\begin{enumerate}
\item A so-called \emph{region of interest} is identified in the
  reference image. This is a region of the image where edges will be
  looked for. The region of interest must be selected to be fully
  contained in both images. This step is done by hand, based on some
  \textit{a priori} knowledge, or visual inspection of the images. In
  our case, where the simulated translations amount to a few pixels
  along either or both the X and Y axes (see
  \S~\ref{sec:simulation-runs}), the region of interest is the whole
  reference image, minus a few pixels around the edges.  In the case
  of two partially overlapping images of the same region one would
  choose the intersection of the two images.
\item Features are identified in the comparison image with an edge
  detection algorithm. Pixels identified by the edge detection
  algorithm are part of the \textit{reference pattern}.
\item The edge detection algorithm is run on the comparison image.
  This results in a second collection of pixels, the \emph{search
    pattern}.
\item The algorithm now overlays the reference pattern on the search
  pattern. The reference pattern is stepped over the search pattern in an
  attempt to maximize the number of overlapping pixels. In doing so
  the algorithm is allowed to reduce the number of pixels in the
  reference pattern. The maximum fraction of the search pattern that can be
  discarded in the process can be set by the user. In our
  application the reference and search patterns can differ
  vastly. Therefore we have allowed the algorithm to throw away up to
  $70\,\%$ of the pixels. In trying to maximize the overlap between
  the two patterns, the algorithm can be instructed to allow for
  a rotation and a scaling factor.
\item The algorithm reports the recovered translations and the
  fraction of the pixels in the reference pattern that was used to
  find a match. The latter is called the \textit{score}. Within the
  parameters given by the user, the algorithm always chooses the match
  with the highest score.
\end{enumerate}

\subsubsection{The Meaning of the Score}
\label{sec:meaning-score}

The HALCON score is the normalized sum of the cross product between
the vectors describing the position of the pixels in the reference
pattern and those describing the pixels in the search pattern. If the
two patterns are identical, it is clear that the score will be equal
to one. When pixels have to be dropped from the reference pattern, the
score will decrease.

In the actual algorithm, the sum of the cross products of the pixels
used in the match is slightly modified to take into account the
possibility of non-linear changes in the illumination conditions,
either locally, around certain features, or globally, across the
entire image~\cite{espacenet}.

It is tempting to interpret the HALCON score as a quality factor for
the goodness of the translation parameters found. However this would
be wrong on two counts.

First of all, it is clear that often only a subset of the pattern to
be looked for is to be found in the search pattern. (Refer back to
the examples shown in figure~\ref{fig:sample-images}.) In this case the
search algorithm must discard some of the pixels in the reference
pattern in order to find a good matching sub-pattern. How many pixels
are left in the sub-pattern has nothing to do with whether the match
is good or not.

Second, the notion of \textit{goodness of match} implies that the
matched pattern can be compared to an expected result, or true
pattern, or that the algorithm proceeds through the optimization of an
objective function. But the only measure of how well the algorithm has
performed, is how close the recovered translation is to the values
injected in the simulation. This means that the accuracy of the
algorithm can only be judged through an extensive set of Monte Carlo
simulations. Only by repeatedly comparing two images in multiple
realizations of the same detection and matching process, is it
possible to gauge the statistical errors in the results, and therefore
establish to what accuracy and under what conditions the algorithm can
be effectively employed.

\section{Approach}
\label{sec:approach}

The following steps were identified.

\begin{enumerate}
\item Render a digital elevation model of the surface of a planet, by
  choosing the position of the Sun and of the spacecraft. We have used
  \textbf{povray}~\cite{povray} for this task.
\item Create two images, the reference image and the comparison image.
  The latter is possibly translated along one or both of the image axes.
\item Convert the rendered images to instrument count images
  of a given signal-to-noise ratio.
\item Recover translation parameters between the two images using a
  shape-based pattern matching algorithm.
\item Study the accuracy with which the parameters are
  recovered, and derive information on the range of
  illumination conditions for which the parameters can be
  successfully recovered.
\end{enumerate}

\section{Digital Elevation Models}
\label{sec:digit-elev-models}

\begin{figure}[htbp]
  \centering
  \includegraphics[width=8cm]{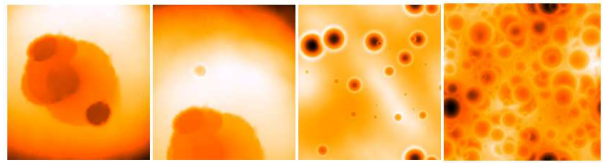}
  \caption{The four digital elevation models used in this study.  From
    left: The Olympus Mons caldera; a bowl-shaped crater close to the
    Olympus Mons caldera; a synthetic landscape with several
    bowl-shaped craters; another synthetic landscape containing
    approximately 5000 craters. (An image of Mercury taken from an
    height of 400\,km might contain a few thousands of craters with a
    diameter larger than a few tens of
    meters.) Darker colors represent lower
    elevations.}
\label{fig:four_dems}
\end{figure}

Four digital elevation models have been used in this
study. These are shown in
figure~\ref{fig:four_dems}\footnote{The Olympus Mons digital elevation model was kindly
  provided to us by the Mars Express Team.}.

\section{Simulation Runs}
\label{sec:simulation-runs}

After some preliminary simulations used to determine a useful sampling
scheme of the parameter space, the bulk of the simulations were
carried out with the following parameters values:

\begin{itemize}
\item Translations: 100 meter in X, in Y, and in X and Y.
\item Sun elevation angles: 10\deg, 30\deg, 60\deg, 90\deg.
\item Nominal spacecraft height 1500\,km\,\footnote{The actual height
    of the camera above the surface is not important for the results
    of this study, at least as long as the images recorded from
    different heights show the same level of detail.}.
\item Sun azimuth angle: several (the same azimuth angle for reference
  and comparison images). For one model a difference in azimuth of
  $30^\circ$ between the two images was introduced.
\item Four digital elevation models rendered with a signal-to-noise
  ratio of 50. The signal-to-noise ratio is determined when the Sun is
  at the zenith.
\end{itemize}

\section{Results}
\label{sec:results}

For each digital elevation model used, several thousand data points
have been calculated. Each data point refers to a particular
combination of Sun elevation and azimuth for the reference and
comparison images, and a translation along one or both of the image
axes. For each combination of parameters, the same number of
simulation runs (ten) was carried out.

In the following we use $\Delta_x$ and $\Delta_y$ to indicate the
difference between the amplitude of the translation recovered by the
algorithm and the amplitude of the translation used in the simulation.
Therefore the expectation value of $\Delta_x$ and $\Delta_y$ is always
0, and the width of their distributions is a measure of the
statistical error in the reconstruction.

In table~\ref{tab:stats} we give a summary of how successful the
algorithm has been. For each digital elevation model we give the
number of realizations (all Sun angles and all translations), how many
times the algorithm failed to return a match, and how many times the
returned result was more than 2 pixels away from the expected result.
The latter figure has no special meaning, but is meant to give an idea
of the global behavior of the algorithm.

\begin{table}[htbp]
  \centering
  \begin{tabular}{|c|c|c|c|}
    \hline
    { DEM }& { Total } & { No match } & {
      $\Delta_{x} >2$ or $\Delta_{y} >2$  } \\
    \hline
    \hline
    A & 9878 & $2.8\%$ & $2.3\%$ \\
    B & 7300 & $3.2\%$ & $4.2\%$ \\
    C & 6254 & $0\%$   & $12.6\%$\\
    \hline
  \end{tabular}
  \caption{Global success statistics for the simulation runs. For each
    digital elevation model (DEM) the following data are reported: the total number of
    independent realizations; the fraction of realizations for which the
    algorithm was not able to find any match; the fraction of realizations
    for which either $|\Delta_x|$  or $|\Delta_y|$ were larger than
    two pixels. DEM keys: A = Olympus Mons, B = bowl crater, C = synthetic.}
  \label{tab:stats}
\end{table}

One thing is immediately apparent: for the synthetic digital elevation
model the algorithm always returned a match, but a larger fraction of
the returned answers was significantly wrong. Because the synthetic
model is significantly more regular than the other two --- in
particular the craters are identical but for a scale factor, the
algorithm has an easier job at finding some matching pattern, although
relatively more often the pattern found is not the good one.

Based on these observations, we present the results for the synthetic
model separately. However, we will be able to show that the same
conclusions on the accuracy of the algorithm can be reached for all
digital elevation models by applying the same selection criteria on
the illumination conditions.

In the following sections we use the following notation:

\begin{itemize}
\item $\theta_{cut}$ refers to the following selection:  $\theta_{sun} >
    10^\circ$ and $\theta_{sun} \ne 90^\circ$\,, where $\theta_{sun}$
    is the Sun elevation angle in the reference or the comparison image.
\item $\phi_{cut}$ refers to the following selection:  $|\phi_{sun} -
  n\times 90^\circ| > 20$ for $n \in {0, 1, 2, 3, 4}$\,, where $\phi_{sun}$
  is the Sun azimuth --- this is the same in both the reference and the
  comparison images.
\end{itemize}

\subsection{The Olympus Mons Models}
\label{sec:olympus-mons-models}

\begin{figure}[tbp]
  \centering
    \includegraphics[width=5cm]{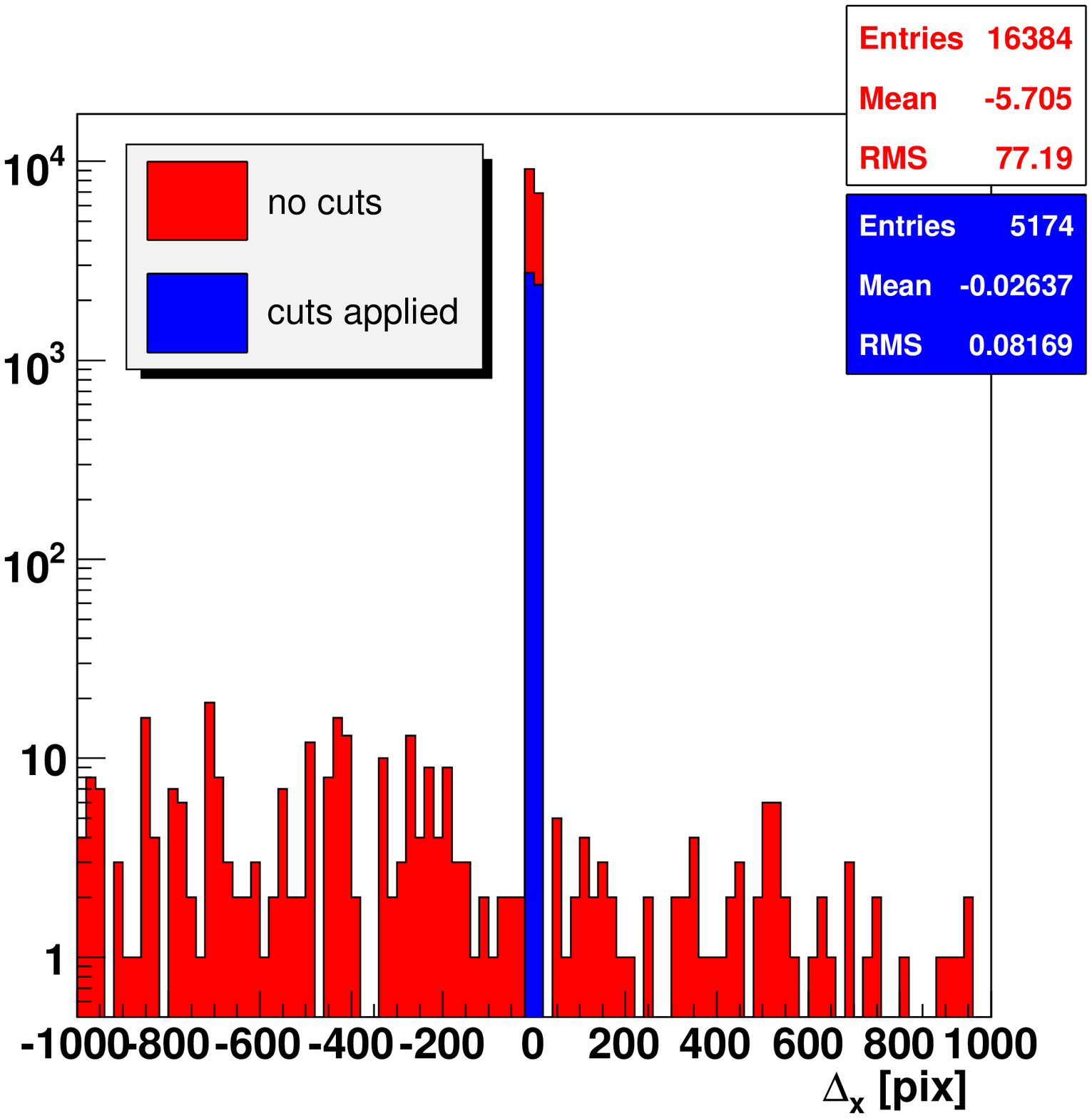}
  \centering
    \includegraphics[width=5cm]{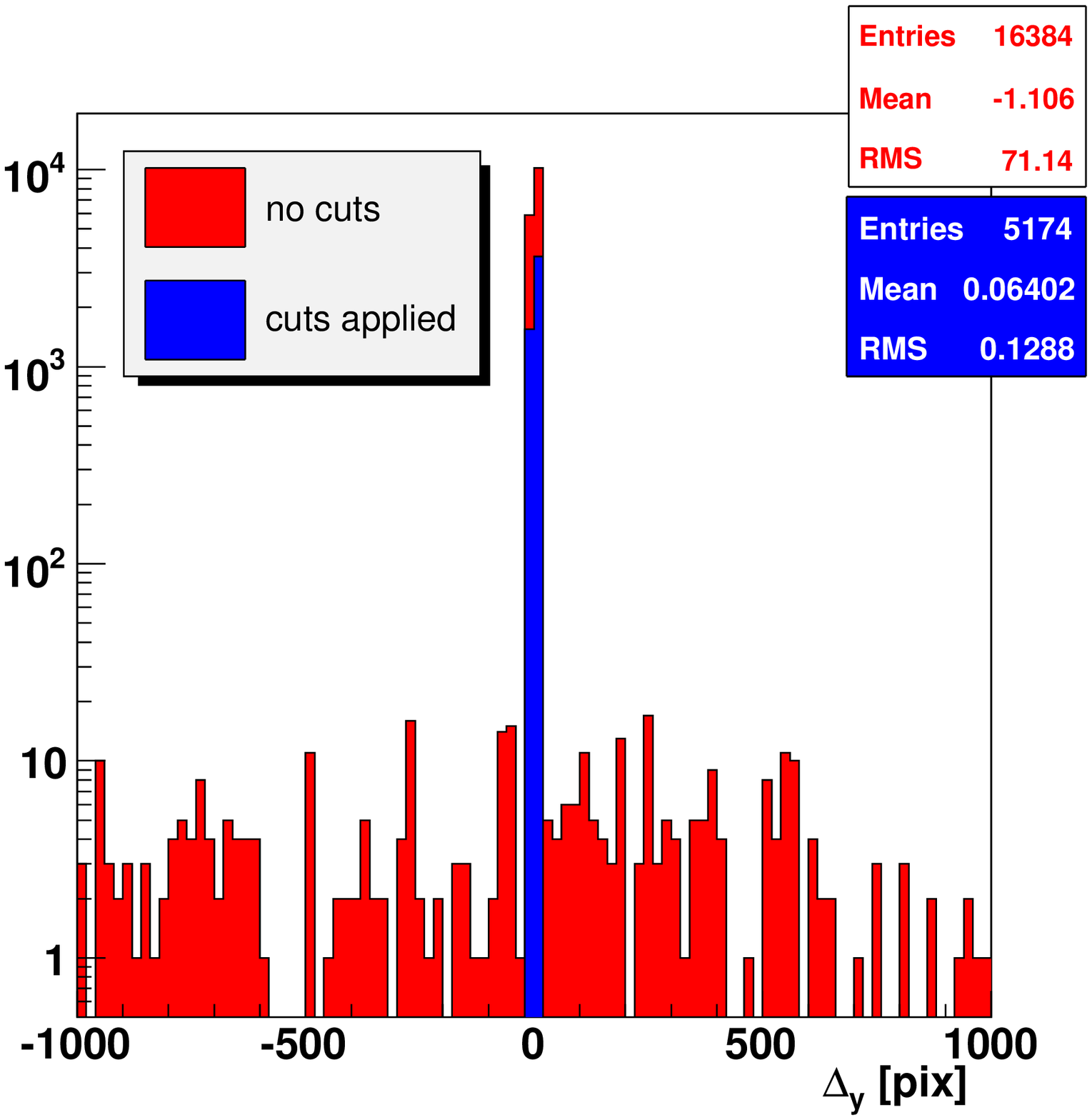}
  
  \caption{The effect of the Sun elevation cut on the distribution of
    $\Delta_x$ (top) and $\Delta_y$ (bottom) for the Olympus Mons data.}
  \label{fig:om-score-deltas-sunel}
\end{figure}

Most of the high-deviation points come from images where the Sun
elevation is equal to $90^\circ$, or images where the Sun elevation is
lower than $10^\circ$ ($\theta_{cut}$).  This is shown in
figure~\ref{fig:om-score-deltas-sunel}.

\begin{figure}[htbp]
  \centering
    \includegraphics[width=5cm]{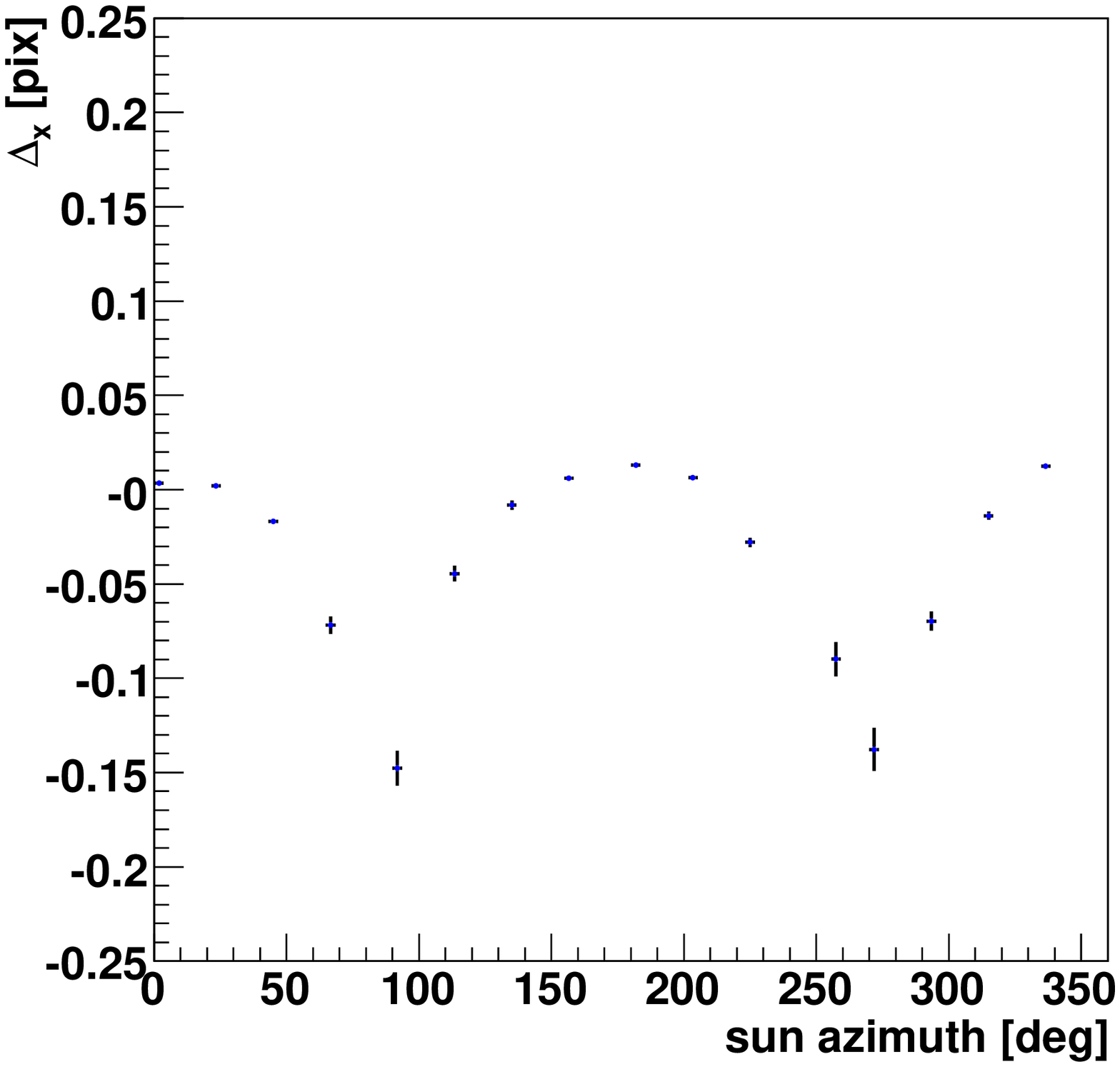}
    \includegraphics[width=5cm]{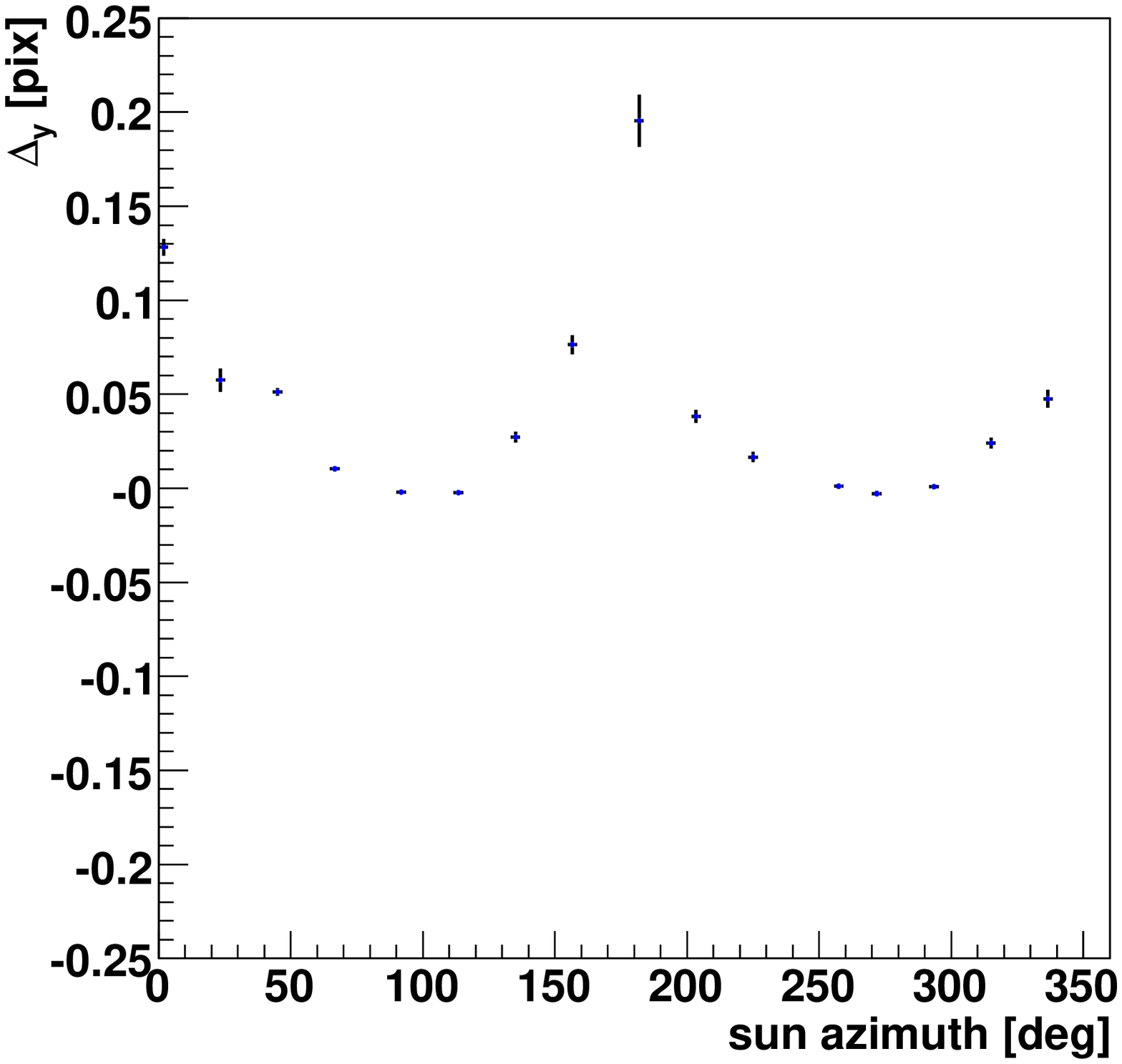}
  \caption{The average $\Delta_x$ and $\Delta_y$ versus the Sun
    azimuth for the Olympus Mons data. The error bar on each point
    represents the root-mean-square.}
  \label{fig:om-deltas-sunaz}
\end{figure}

In figure~\ref{fig:om-deltas-sunaz} we plot the data with
$\theta_{cut}$ applied versus the Sun azimuth. We observe that the
mean of $\Delta_x$ and $\Delta_y$ vary with $\phi_{sun}$ in a
quasi-periodic fashion. What we observe is that the deviation is larger
when the Sun azimuth is orthogonal to one of the image axes. Namely,
the largest deviations for $\Delta_x$ are observed when $\phi_{sun}
\approx 90^\circ$ or $270^\circ$, whereas the largest deviations for
$\Delta_y$ are observed when $\phi_{sun} \approx 0^\circ$ and
$180^\circ$\,. The direction defined by the Sun azimuth appears to be
a preferential direction: translations along this direction can be
more accurately recovered, because the features on the terrain create
sharper shadows along the direction to the Sun.

\begin{figure}[htbp]
  \centering
    \includegraphics[width=5cm]{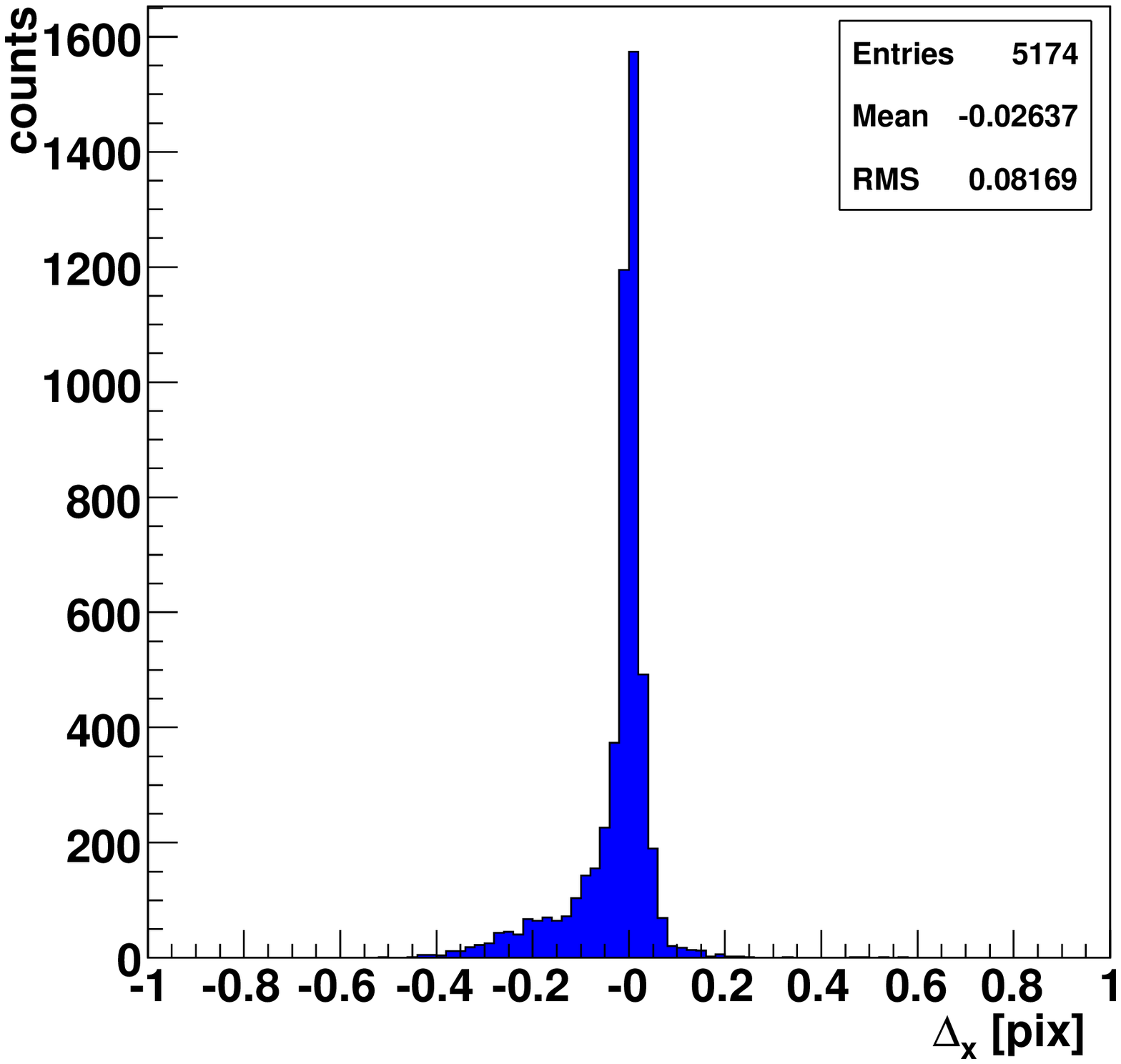}
    \includegraphics[width=5cm]{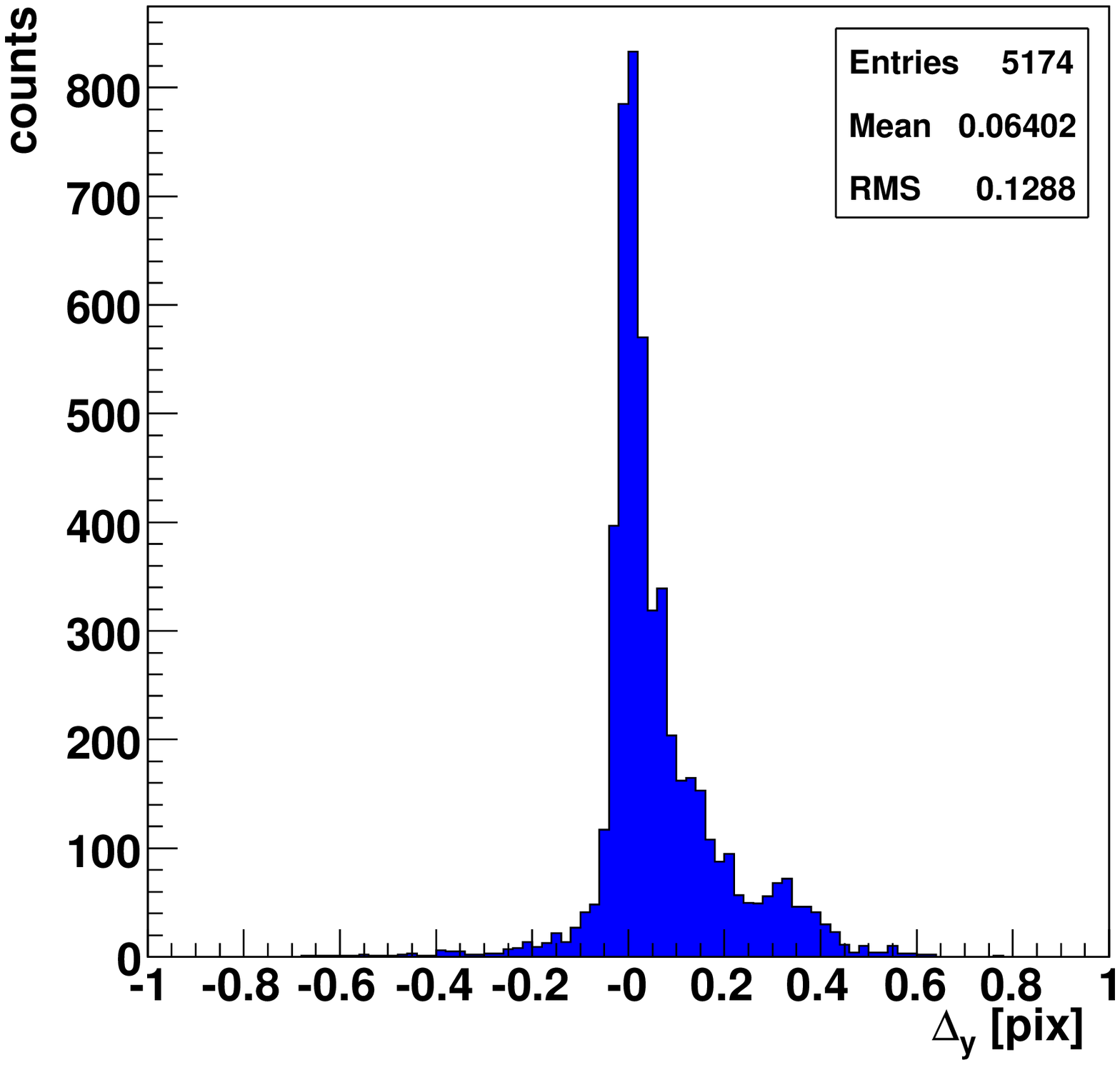}
  \caption{The distribution of $\Delta_x$ and $\Delta_y$ for the
    Olympus Mons data, once both the Sun cuts are applied.}
  \label{fig:om-histo-deltas}
\end{figure}

Based on the data in figure~\ref{fig:om-deltas-sunaz} we can devise a
selection criterion for the Sun azimuth, so that translations along
both axes can be recovered with comparable accuracy. The selection
criterion is that the Sun azimuth must be more than $20^\circ$ away
from both image axes. The distributions of $\Delta_x$ and
$\Delta_y$ when both $\theta_{cut}$ and $\phi_{cut}$ are applied are
shown in figure~\ref{fig:om-histo-deltas}.

Figure~\ref{fig:om-histo-deltas} represents the end point of our
analysis. We observe that the two distributions are centered on 0, and
have a width of $\approx$\,0.1\,pixel root-mean-square. The two
distributions have a tail in the direction of the translation applied
in the simulations (-100\,m along the X axis, and +100\,m along the Y
axis). The nature of this slight asymmetry in not understood at present.

\subsubsection{Changing the Sun Azimuth}
\label{sec:changing-sun-azimuth}

\begin{figure}[htbp]
  \centering
    \includegraphics[width=5cm]{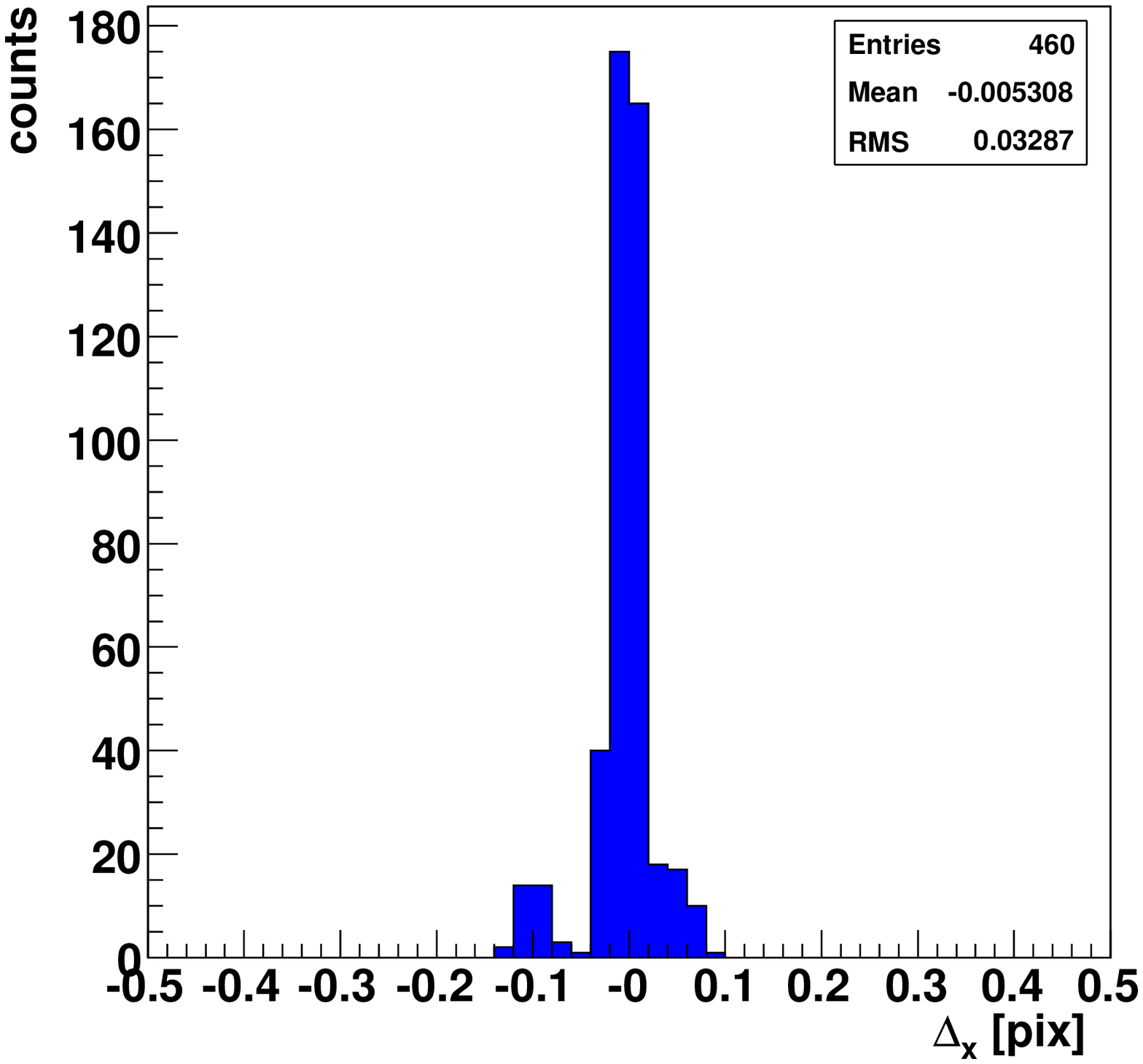}
    \includegraphics[width=5cm]{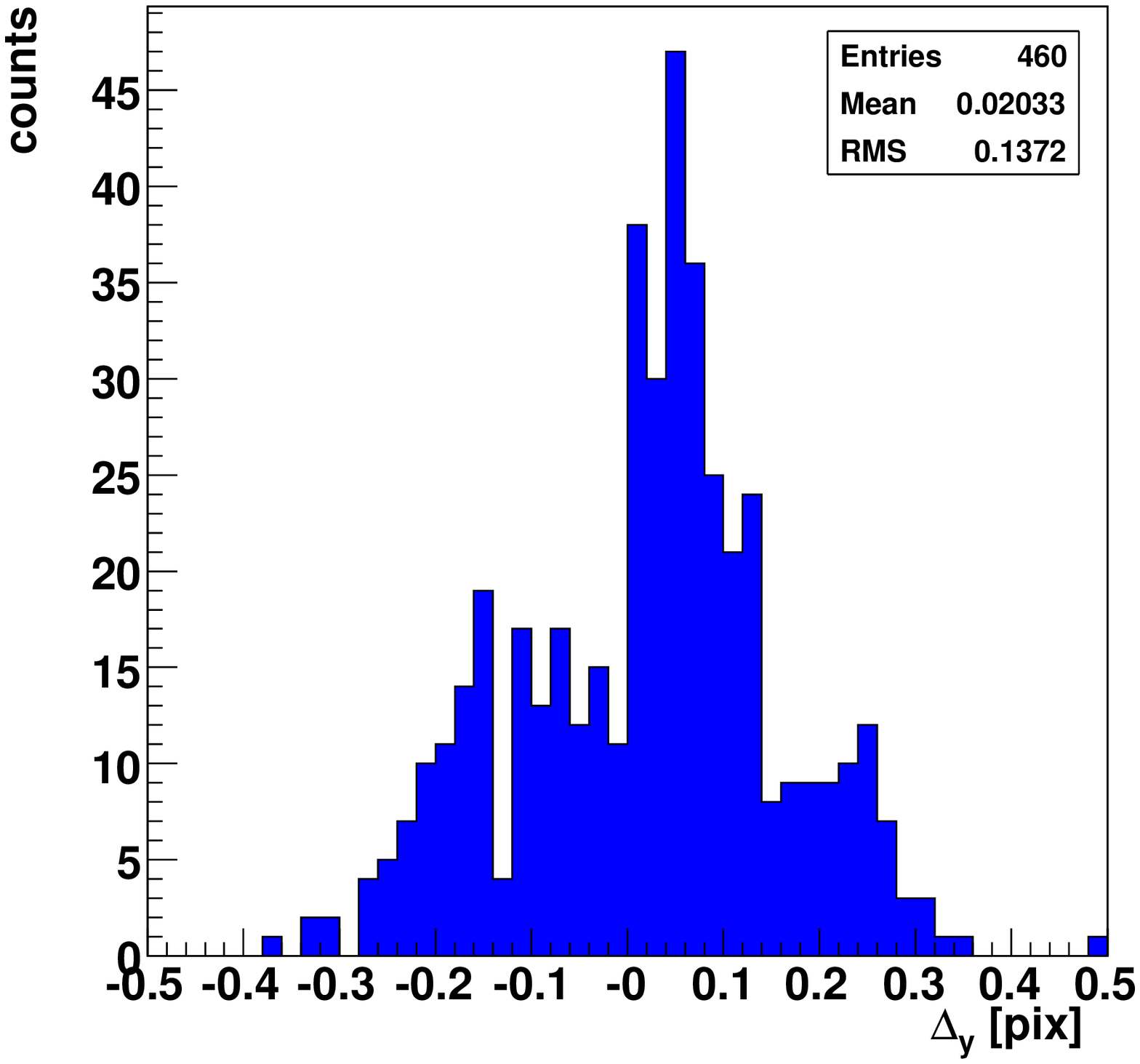}
   \caption{The distribution of $\Delta_x$ (top) and $\Delta_y$
     (bottom) for the
    Olympus Mons data for date where the Sun azimuth of the comparison
  and reference images differ by $30^\circ$.}\label{fig:daz}
\end{figure}

The bulk of the simulation runs was carried out with the same Sun
azimuth for both the reference and comparison images. We however also
made a set of simulation runs where the azimuth of the Sun in the
comparison image was $30^\circ$ away from the azimuth used in the
reference image; only a translation of 100 meters along the X axis was
applied. The results are shown in figure~\ref{fig:daz}. Even in this
case the algorithm is able to recover the injected translation with
an accuracy of $\approx$\,0.1\,pixel root-mean-square.

\subsection{The Synthetic Model}
\label{sec:synthetic-model}

As already hinted to, the results based on the synthetic digital
elevation model give a slightly different picture, although the main
conclusions do not change.

The $\theta_{cut}$ criterion is still effective in rejecting data
points that return a large deviation from the expectation.

\begin{figure}[htbp]
  \centering
    \includegraphics[width=5cm]{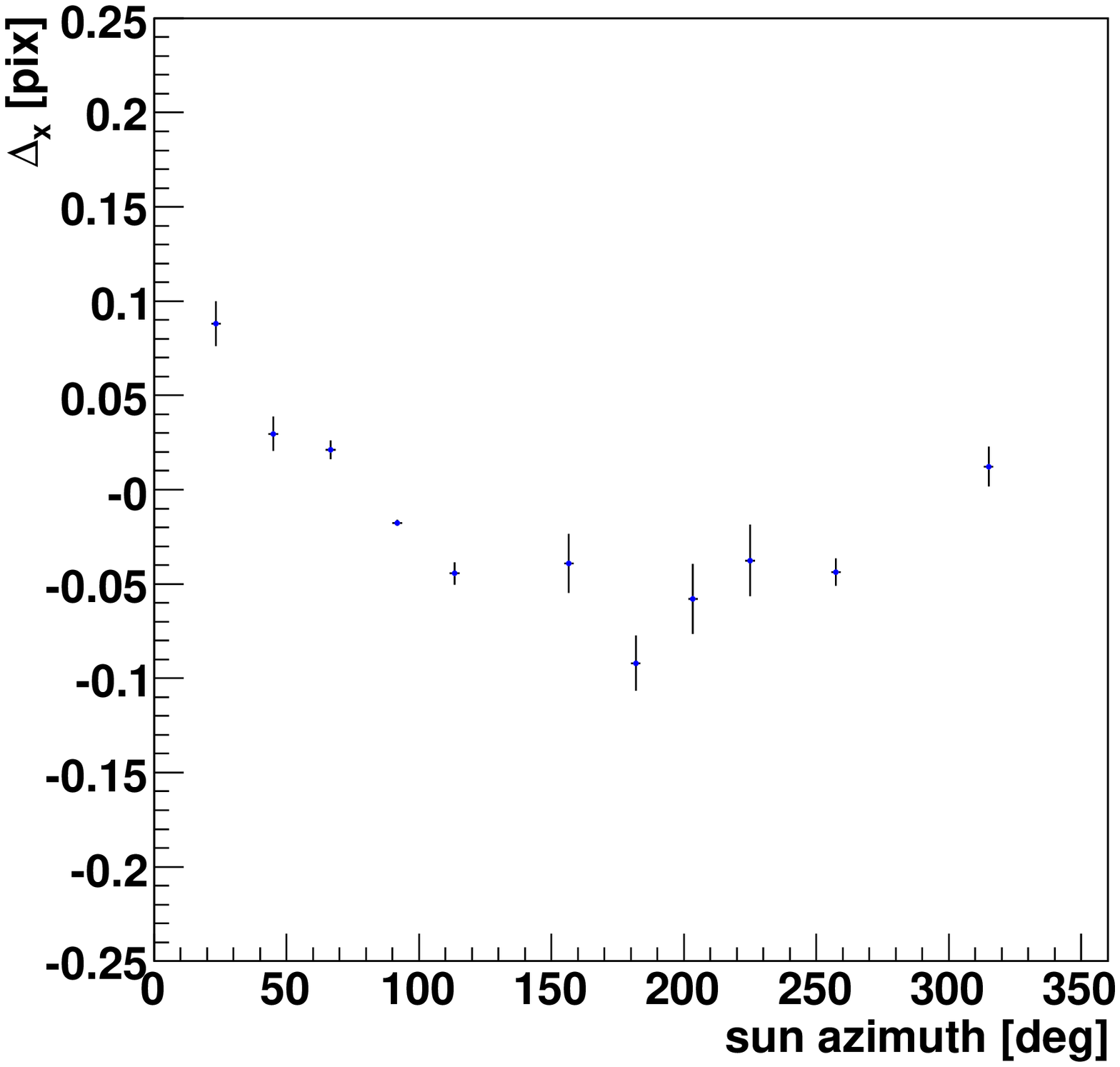}
    \includegraphics[width=5cm]{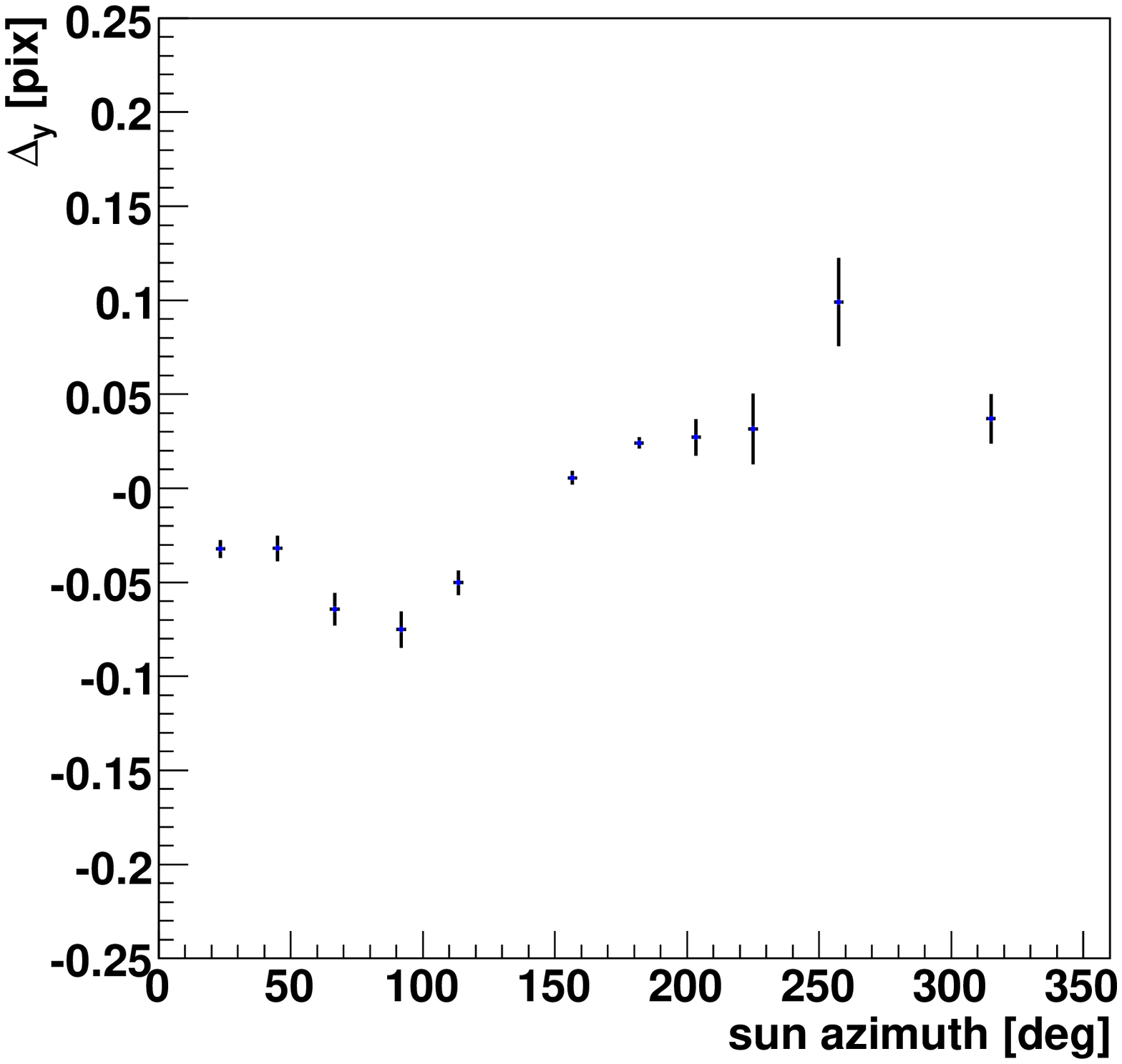}

  \caption{The average $\Delta_x$ (top) and $\Delta_y$ (bottom) versus the Sun
    azimuth for the Synthetic Model data. The error bar on each point
    represents the root-mean-square.}
  \label{fig:sc-deltas-sunaz}
\end{figure}

A point of discrepancy with respect to the Olympus Mons data is the
behavior of the recovered translations as a function of Sun azimuth.
Figure~\ref{fig:sc-deltas-sunaz} shows that the effect observed for
the Olympus Mons models is almost not observed here. After the
$\theta_{cut}$ criterion is applied, any remaining offset is smaller
than 0.05\,pixel.

\begin{figure}[htbp]
  \centering
    \includegraphics[width=5cm]{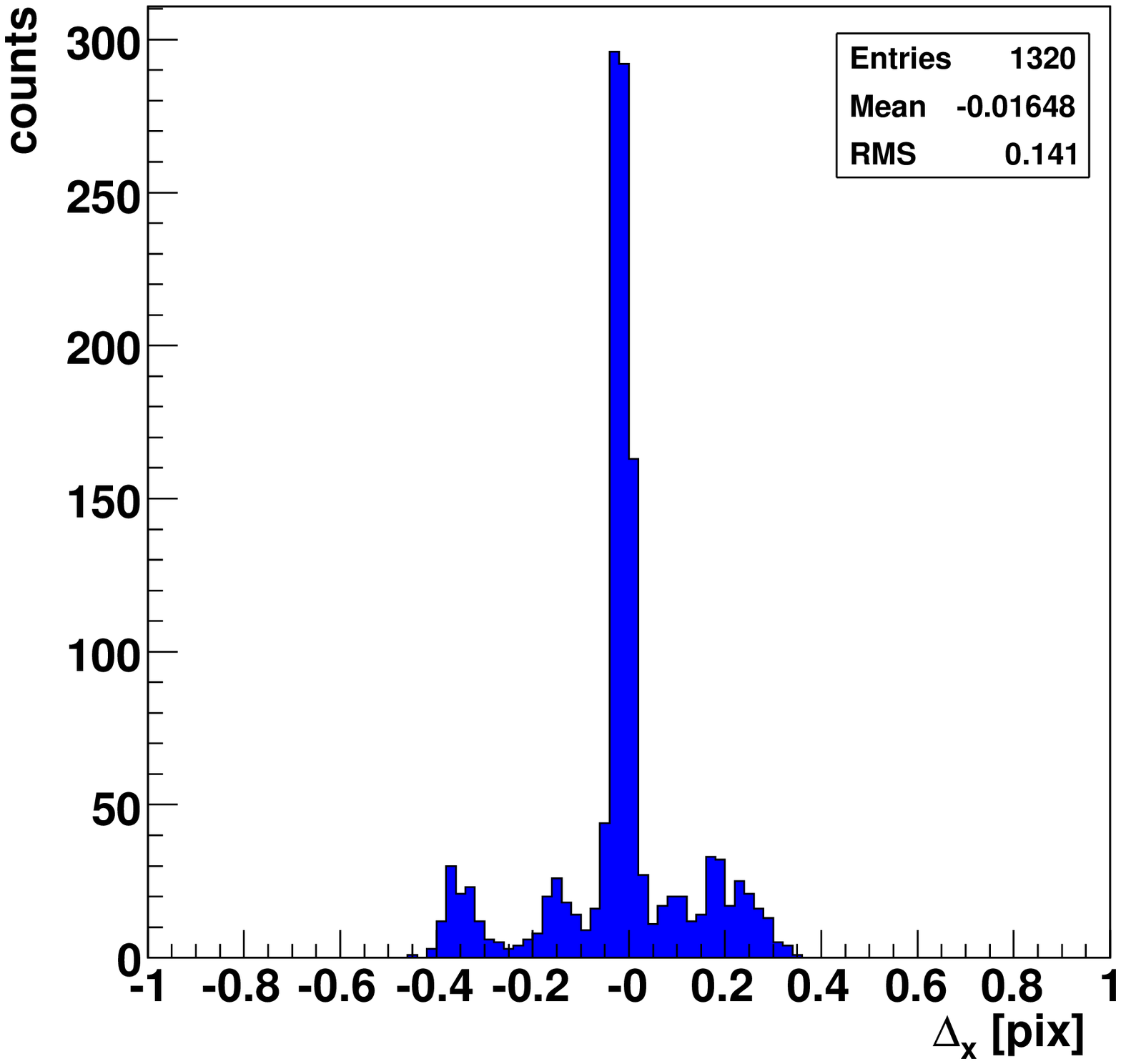}
    \includegraphics[width=5cm]{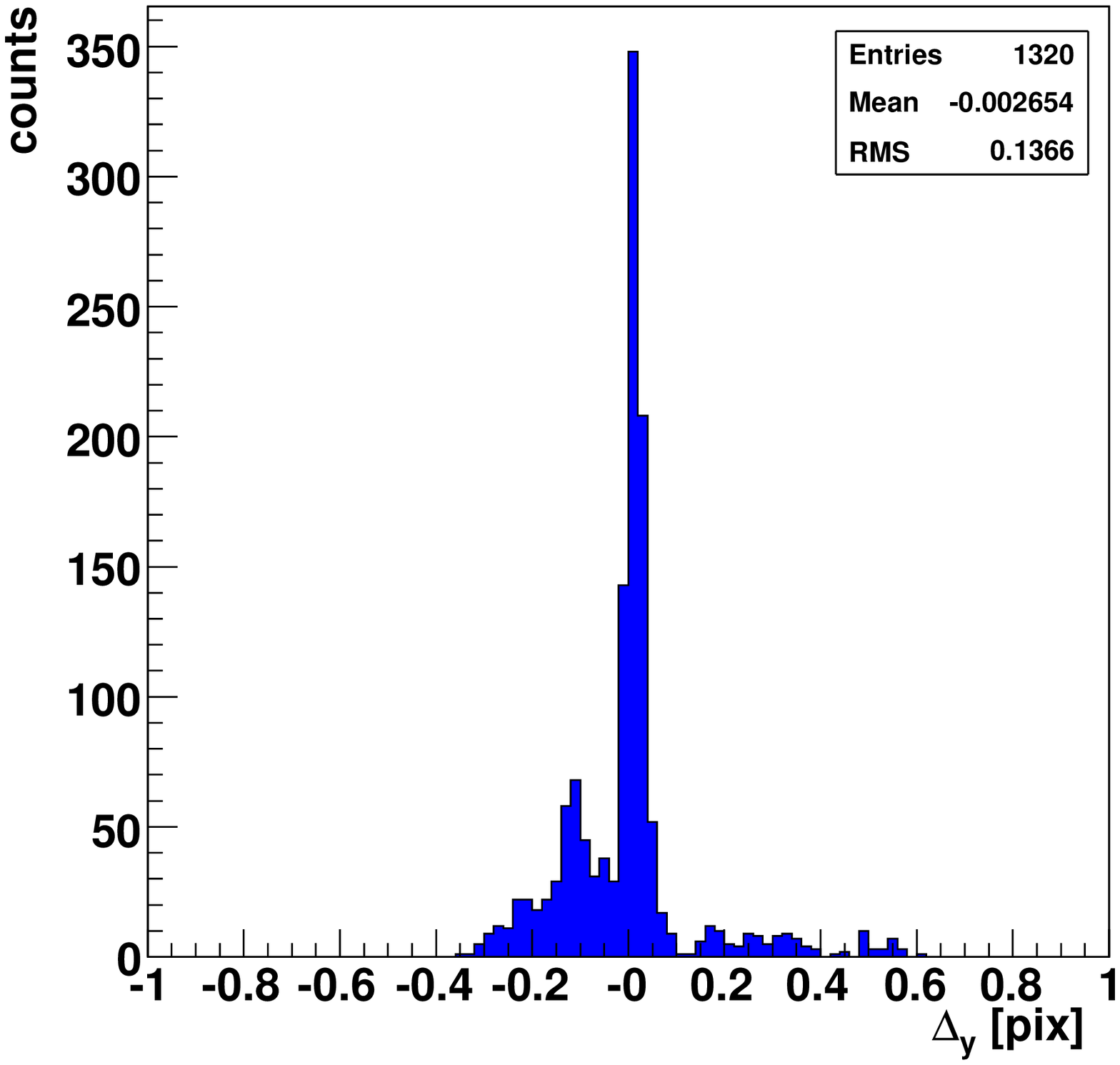}
  
  \caption{The distribution of $\Delta_x$ (top) and $\Delta_y$
    (bottom) for the
    Synthetic Model data, once both the Sun cuts are applied.}
  \label{fig:sc-histo-deltas}
\end{figure}

Finally, figure~\ref{fig:sc-histo-deltas} shows the distribution for
$\Delta_x$ and $\Delta_y$. Again, the translation is recovered with an
accuracy of  $\approx$\,0.1\,pixel root-mean-square, but the details of
the distributions differ from what was observed before.

\section{Conclusions}
\label{sec:conclusions}

We have performed a study of the accuracy with which a shape-based
pattern matching algorithm can identify translations between remote
sensing images of the same planetary features. We have applied the
algorithms in a Monte Carlo fashion to digital elevation models (both
real and synthetic) in order to investigate the statistical performance
of the procedure.

We find that for a broad range of illumination conditions translations
between images can be recovered with an accuracy of 0.1\,pixel
\textit{r.m.s}.

The algorithm performs best for translations along the projected
direction to the Sun on the image plane. This study shows that
translations along both image axes at the same time can be recovered
with the same accuracy of 0.1\,pixel as long as the projected
direction to the Sun lies more than $\approx 20^\circ$ away from the
same image axes.

Finally, this study demonstrates that the images to be compared need
not be taken under the very same illumination conditions in order to
be effectively matched. For a given Sun azimuth, any pair of images
taken with Sun elevation angles larger than $10^\circ$ can be used;
images taken when the Sun is at the zenith must also be avoided. The
range of useful illumination conditions is further broadened because
this study concludes that differences in Sun azimuth of at least
$30^\circ$ do not affect the accuracy of the matching algorithm.

The error contributed by the matching algorithm is but one of the
several error contributions to be taken into account during the
analysis of the data pertaining to the measurement of the possible
libration of the surface of Mercury. This study shows that the
accuracy of the pattern matching algorithm is not a limiting factor in
the ultimate accuracy of the libration experiment aboard the
BepiColombo mission to Mercury.

\section*{Acknowledgments}
\label{sec:acknowledgements}

This study was carried out under ESA contract ESTEC 18624.

\raggedright{}

\bibliography{./bibliography}
\bibliographystyle{esa}

\end{document}